\documentclass[final,1p,times]{elsarticle}
\usepackage{amssymb,amsmath}
\biboptions{compress}

\journal{Nuclear Physics B}

\begin{document}

\begin{frontmatter}

\title{Non-commutative geometry and thermodynamics of the Schwarzschild-AdS black hole}

\author{Slimane Zaim}
\author{Fatma Zohra Bara}
\author{Mohamed Aimen Larbi}
\affiliation{organization={Department of Physics, Faculty of Matter Sciences, University of Batna-1},
            city={Batna},
            postcode={05000}, 
            country={Algeria}}

\begin{abstract}
We investigate the thermodynamic properties of a Schwarzschild anti-de Sitter (AdS) black hole within the framework of noncommutative geometry. We derive the relevant thermodynamic quantities and show that they depend sensitively on the noncommutativity parameter, denoted by $\Theta$, while consistently satisfying the first law of black hole thermodynamics. A detailed stability analysis reveals the existence of a phase transition at a critical point. Furthermore, the thermodynamic behavior exhibits a close analogy with that of a Van der Waals fluid, where noncommutative effects introduce corrections to the black hole temperature. Our findings suggest that the parameter $\Theta$ is naturally of the order of the Planck length ($\ell_p$) and can be interpreted as an additional thermodynamic variable characterizing the system.
\end{abstract}


\end{frontmatter}

\section{Introduction}

Since the pioneering work of Hawking and Page in 1983~\cite{Hawking:1982dh}, who first discovered the phase transition in Schwarzschild--AdS (SAdS) black holes, black hole thermodynamics has attracted considerable attention and has led to many profound results. By identifying the black hole mass and cosmological constant with enthalpy and pressure, respectively, the first law of AdS black hole thermodynamics was established~\cite{Kastor:2009wy}, enabling a deeper understanding of the thermodynamic structure of AdS black holes.

The study of black holes in anti-de Sitter (AdS) spacetime is particularly compelling because of their rich thermodynamic behavior, including stability properties and phase structure~\cite{Hubeny:2014bla, Dolan:2010ha}. In this context, the black hole mass has been interpreted as the spacetime enthalpy~\cite{Kastor:2009wy, Dolan:2010ha}. Over the past decade, this perspective has been extended by promoting the cosmological constant to a thermodynamic variable associated with pressure, with its conjugate quantity identified as thermodynamic volume. This framework, known as extended phase space thermodynamics (EPST)~\cite{Dolan:2011xt,Dolan:2011jm,Kubiznak:2016qmn,Kubiznak:2012wp}, has led to the discovery of a variety of intriguing phenomena, including phase transitions between small and large black holes and analogies with Van der Waals fluids~\cite{Kubiznak:2012wp,Cai:2013qga,Hendi:2012um,Dehghani:2014caa,Liu:2016uyd}.

More recently, Gao and Zhao~\cite{Zeyuan:2021uol,Gao:2021xtt} highlighted several conceptual issues within the EPST framework, such as the non-homogeneous character of the Smarr relation and the interpretation of black hole mass as enthalpy. Motivated by these concerns, they proposed an alternative framework, termed restricted phase space thermodynamics (RPST), in which the central charge and its conjugate chemical potential are introduced as fundamental thermodynamic variables~\cite{Zeyuan:2021uol,Gao:2021xtt,Visser:2021eqk}. In RPST, the cosmological constant is held fixed while Newton's constant is allowed to vary, replacing the pressure--volume pair with the central charge--chemical potential pair. In this setup, the Smarr relation becomes first-order homogeneous in the extensive variables, and the black hole mass is interpreted as internal energy. This formalism has since inspired a range of interesting developments~\cite{Sadeghi:2022jlz,Du:2022ckz}.

Within the RPST approach, thermodynamic quantities of AdS black holes have also been related to those of the dual conformal field theory (CFT), and a holographic first law has been formulated in a way consistent with black-hole extended thermodynamics~\cite{Visser:2021eqk,Maldacena:1997re}. These developments suggest that the formalism may offer deeper insight into the interplay between gravity and quantum physics. More broadly, the thermodynamic properties of AdS black holes are intimately tied to their horizon structure, hinting at a profound connection between black hole physics and quantum phenomena through the quantization of spacetime itself. In this regard, noncommutative (NC) geometry provides a natural framework for describing spacetime quantization~\cite{Doplicher:1994tu,Frob:2022ciq,Seiberg:1999vs}.

NC  geometry has proven to be particularly significant, as it provides a consistent deformation of spacetime that allows the investigation of quantum gravitational corrections to classical thermodynamic quantities. NC geometry, inspired by string theory~\cite{Seiberg:1999vs,Witten:1985cc,Chu:1998qz}, treats spacetime coordinates as noncommuting operators, thus introducing a natural ultraviolet cutoff.

By analogy with the commutation relation between position and momentum in quantum mechanics, NC geometry postulates the following commutator between spacetime coordinates
\begin{equation}
[\hat{x}^{\mu}, \hat{x}^{\nu}] = i \Theta^{\mu \nu},\label{eq1}
\end{equation}
where $\overline{\Theta}^{\mu \nu}$ is an antisymmetric constant tensor. The parameter $\Theta$ has dimensions of $(\text{length})^{2}$ and characterizes the minimal spacetime scale, with $\sqrt{\Theta}$ typically assumed to be of the order of the Planck length $\ell_{p}$~\cite{Doplicher:1994tu,Frob:2022ciq,Seiberg:1999vs}. The Bopp shift gives the relation between the commutative and noncommutative coordinate systems
\begin{equation}
\hat{x}^{\mu} = x^{\mu} - \frac{1}{2}\overline{\Theta}^{\mu \nu} p_{\nu},\label{eq2}
\end{equation}
where $\overline{\Theta}^{\mu \nu} = \Theta^{\mu \nu}/\hbar$. In NC spacetime, the usual product of functions is replaced by the Moyal (or star) product
\begin{align}
(f \ast g)(x) &= f(x) \, e^{\frac{i}{2}\Theta^{\mu \nu}\overleftarrow{\partial_{\mu}} \partial_{\nu}} g(y) \big|_{y = x}\notag \\
&= f(x)g(x) + \frac{i}{2}\Theta^{\mu \nu} \partial_{\mu}f(x)\, \partial_{\nu}g(x) + \mathcal{O}(\Theta^{2}).\label{eq3}
\end{align}

Numerous studies have explored the thermodynamic properties of black holes within the framework of NC geometry~
\cite{Heidari:2023egu,Filho:2024zxx,Touati:2022zbm,Hamil:2020cwu,Hamil:2021ilv,Hamil:2021asg,Hamil:2022bpd,Chen:2022ngd,Chen:2022dap,Touati:2023gwv,Touati:2023joy,Nozari:2008rc,Campos:2021sff,Banerjee:2008gc,Sharif:2011ja,Alavi:2009tn,Nozari:2006bi,
Nicolini:2005vd,Wang:2024jlj,Singh:2020xju}. In Refs.~\cite{Nozari:2006bi,Nicolini:2005vd}, NC corrections to the Schwarzschild black hole, modeled through Gaussian and Lorentzian matter distributions, revealed the emergence of a zero-temperature extremal black hole configuration. These analyses also showed the existence of a finite maximum temperature reached during evaporation, after which the black hole cools toward absolute zero. An important outcome of these studies was the observation that the standard first law of black hole thermodynamics is generally violated.

Similar features were found in Ref.~\cite{Wang:2024jlj} for a NC Schwarzschild--AdS black hole based on a Lorentzian distribution, where violations of the first law were likewise reported. More generally, for static spherically symmetric black holes whose energy-momentum tensor depends explicitly on the black hole mass, the usual form of the first law requires modification~\cite{Wang:2024jlj}, which complicates the definition of thermodynamic quantities such as enthalpy and Gibbs free energy.

In this work, we investigate the thermodynamics of Schwarzschild--AdS black holes in NC geometry using the Bopp shift and the Moyal product approach. We show that the thermodynamic observables acquire a nontrivial dependence on the noncommutativity parameter while consistently satisfying both the first law of thermodynamics and the Smarr relation. In addition, the analysis reveals the existence of a finite maximum black hole temperature, as well as a minimum mass induced by noncommutativity below which black hole formation is not possible. Our results show that noncommutativity removes the divergent behavior of the temperature, allowing an estimate of $\Theta \simeq 0.1\, \ell_{p}$. We also analyze the heat capacity to discuss the stability of the NC black hole and examine the Helmholtz free energy to explore the associated phase transition.

This paper is organized as follows. In Section~2, we introduce the NC corrections to the AdS--Schwarzschild metric using Bopp’s shift and derive the relation between total mass and horizon radius. Section~3 presents a detailed analysis of the thermodynamic properties of the AdS--Schwarzschild black hole in NC spacetime. Finally, Section~4 summarizes our main results and conclusions. In this article, we use natural units $G=c=k_B=\hbar=1$.

\section{AdS--Schwarzschild black hole in NC spacetime}

The pioneering works of Bekenstein and Hawking established that black holes possess well-defined thermodynamic properties, with the temperature depending on the radius of the event horizon and the entropy proportional to its surface area~\cite{Hawking:1982dh}. These results laid the foundation for treating black holes as thermodynamic systems. The significance of this framework lies in the fact that black hole thermodynamics follows laws that closely resemble those of classical thermodynamics. In this context, NC geometry, viewed as an approach to spacetime quantization, provides a natural setting in which to investigate black hole thermodynamics, offering a valuable avenue for probing fundamental aspects of quantum gravity.

We are interested in the Schwarzschild--AdS black hole, whose line element is given by
\begin{equation}\label{eq4}
ds^{2}=-f(r)\,dt^{2}+\frac{1}{f(r)}\,dr^{2}+r^{2}\left(d\theta^{2}+\sin^{2}\theta\,d\varphi^{2}\right).
\end{equation}
Substituting this metric into the Einstein equation $(R_{\mu\nu}-\frac{1}{2}Rg_{\mu\nu}+\Lambda g_{\mu\nu}=0)$, one obtains
\begin{equation}\label{eq5}
f(r)=\left(1-\frac{2m}{r}-\frac{\Lambda}{3}r^{2}\right),
\end{equation}
where $m$ represents the Schwarzschild mass and $\Lambda$ is the negative cosmological constant.

In the NC space described by the commutator in Eq.~\eqref{eq1}, the line element becomes
\begin{equation}\label{eq6}
ds^{2}=-\left(1-\frac{2M}{\hat{r}}-\frac{\Lambda}{3}\hat{r}^{2}\right)dt^{2}+\frac{d\hat{r}^{2}}{\left(1-\frac{2M}{\hat{r}}-\frac{\Lambda}{3}\hat{r}^{2}\right)}+\hat{r}^{2}\left(d\theta^{2}+\sin^{2}\theta\,d\varphi^{2}\right),
\end{equation}
where $M$ represents the physical ADM mass of the black hole in NC space and $\hat{r}$ is the radial coordinate operator in NC spacetime.

The NC Schwarzschild--AdS spacetime metric $\hat{g}_{\mu \nu}$, up to first order in $\overline{\Theta}$, takes the form~\cite{Larbi:2024rxv}
\begin{equation}\label{eq7}
ds^{2} = -\hat{g}_{tt}(r,\overline{\Theta})\, dt^{2}+ \frac{dr^{2}}{\hat{g}_{tt}(r,\overline{\Theta})}+ \hat{g}_{\theta \theta}(r,\overline{\Theta})\, d\theta^{2}+ \hat{g}_{\varphi \varphi}(r,\overline{\Theta})\, d\varphi^{2}.
\end{equation}
To simplify calculations, we choose the antisymmetric tensor for the noncommutativity parameter as
\begin{equation}\label{eq8}
\left( \Theta^{\mu \nu} \right) =
\begin{pmatrix}
0 & 0 & 0 & 0 \\
0 & 0 & 0 & -\overline{\Theta} \\
0 & 0 & 0 & 0 \\
0 & \overline{\Theta} & 0 & 0
\end{pmatrix}.
\end{equation}
This choice can be achieved by a rotation or a redefinition of coordinates. We consider only spatial noncommutativity, $\Theta^{0i}=0$, to avoid known issues with unitarity and causality~\cite{Gomis:2000zz,Seiberg:2000gc}.

Using Eq.~\eqref{eq8} in the Bopp shift~\eqref{eq2}, the NC radial coordinate becomes
\begin{equation}\label{eq9}
\hat{r}=r-\frac{\overline{\Theta}}{2}L,
\end{equation}
where $L=p_{\varphi}$ is the angular momentum in the commutative case~\cite{Larbi:2024rxv}. Using the Bopp shift~\eqref{eq2} and the definition of the Moyal product~\eqref{eq3}, together with the components of $\Theta^{\mu \nu}$, we obtain the NC AdS--Schwarzschild metric components $\hat{g}_{\mu \nu}$ up to first order in $\overline{\Theta}$
\begin{align}
\hat{g}_{tt} &= \left( 1 - \frac{2M}{r} - \frac{\Lambda}{3} r^{2} \right)+ \frac{\overline{\Theta} L}{r} \left( \frac{M}{r} - \frac{\Lambda}{3} r^{2} \right)+ \mathcal{O}(\overline{\Theta}^{2}), \label{eq10} \\
\hat{g}_{\theta \theta} &= r^{2} + r \overline{\Theta} L+ \mathcal{O}(\overline{\Theta}^{2}), \label{eq11} \\
\hat{g}_{\varphi \varphi} &= r^{2} \sin^{2}\theta+ r \overline{\Theta} L \sin^{2}\theta+ \mathcal{O}(\overline{\Theta}^{2}). \label{eq12}
\end{align}

We assume a corrected event horizon radius, up to first order, as~\cite{Chaichian:2007dr}
\begin{equation}\label{eq13}
\hat{r}_{h}=r_{h}+C \overline{\Theta}.
\end{equation}
Then
\begin{equation}\label{eq14}
\hat{g}_{00}(\hat{r}_{h},\overline{\Theta})=1-\frac{2M}{r_h+C \overline{\Theta}}-\frac{\Lambda}{3}(r_h+C \overline{\Theta})^2+\overline{\Theta} \frac{ML}{r_h^2}-\overline{\Theta} \frac{\Lambda L}{3}r_h=0.
\end{equation}
Keeping terms up to first order in $\overline{\Theta}$, we obtain
\begin{equation}\label{eq15}
\overline{\Theta} L\left(\frac{M}{r_h^2}-\frac{\Lambda}{3}r_h\right)-2C \overline{\Theta}\left(\frac{M}{r_h^2}-\frac{\Lambda}{3}r_h\right)=0,
\end{equation}
which leads to $C=L/2$. Hence, the horizon shift is
\begin{equation}\label{eq16}
\hat{r}_{h}=r_{h}+\frac{\overline{\Theta} L}{2},
\end{equation}
where $r_h=2M+8\Lambda M^3/3$ corresponds to the horizon radius in the commutative case $(\overline{\Theta}=0)$. The free parameter $M$ is the physical ADM mass of the metric tensor~\eqref{eq7},
defined as the value of the total mass function at spatial infinity.

To derive the relation between the total mass and the horizon radius in NC spacetime, we impose $\hat{g}_{tt}(\hat{r}_{h},\overline{\Theta})=0$, leading to
\begin{align}
M&=\frac{r_{h}}{2}\left(1+\frac{\overline{\Theta} L}{r_{h}}\right)-\frac{\Lambda}{6}r_{h}^{3} \left(1+\frac{3\overline{\Theta} L}{r_{h}}\right)\notag\\
&=\frac{r_{h}}{2}\left(1-\frac{\Lambda}{3}r_{h}^{2}\right)+\left(1-\Lambda r_{h}^{2}\right)\frac{\overline{\Theta} L}{2}+\mathcal{O}(\overline{\Theta}^{2}).\label{eq17}
\end{align}
The first term in Eq.~\eqref{eq17}, $\frac{r_{h}}{2}\left(1-\frac{\Lambda}{3}r_{h}^{2}\right)$, represents the black hole mass in AdS spacetime, while the second term, $\left(1-\Lambda r_{h}^{2}\right)\frac{\overline{\Theta} L}{2}$, corresponds to the correction induced by NC geometry. In the limit $\overline{\Theta}\to 0$, the standard commutative AdS black hole mass is recovered
\begin{equation}\label{eq18}
M \to m = \frac{r_{h}}{2} - \frac{\Lambda}{6}r_{h}^{3}.
\end{equation}
\begin{figure}[ht]
\centering
\includegraphics[width=0.8\linewidth]{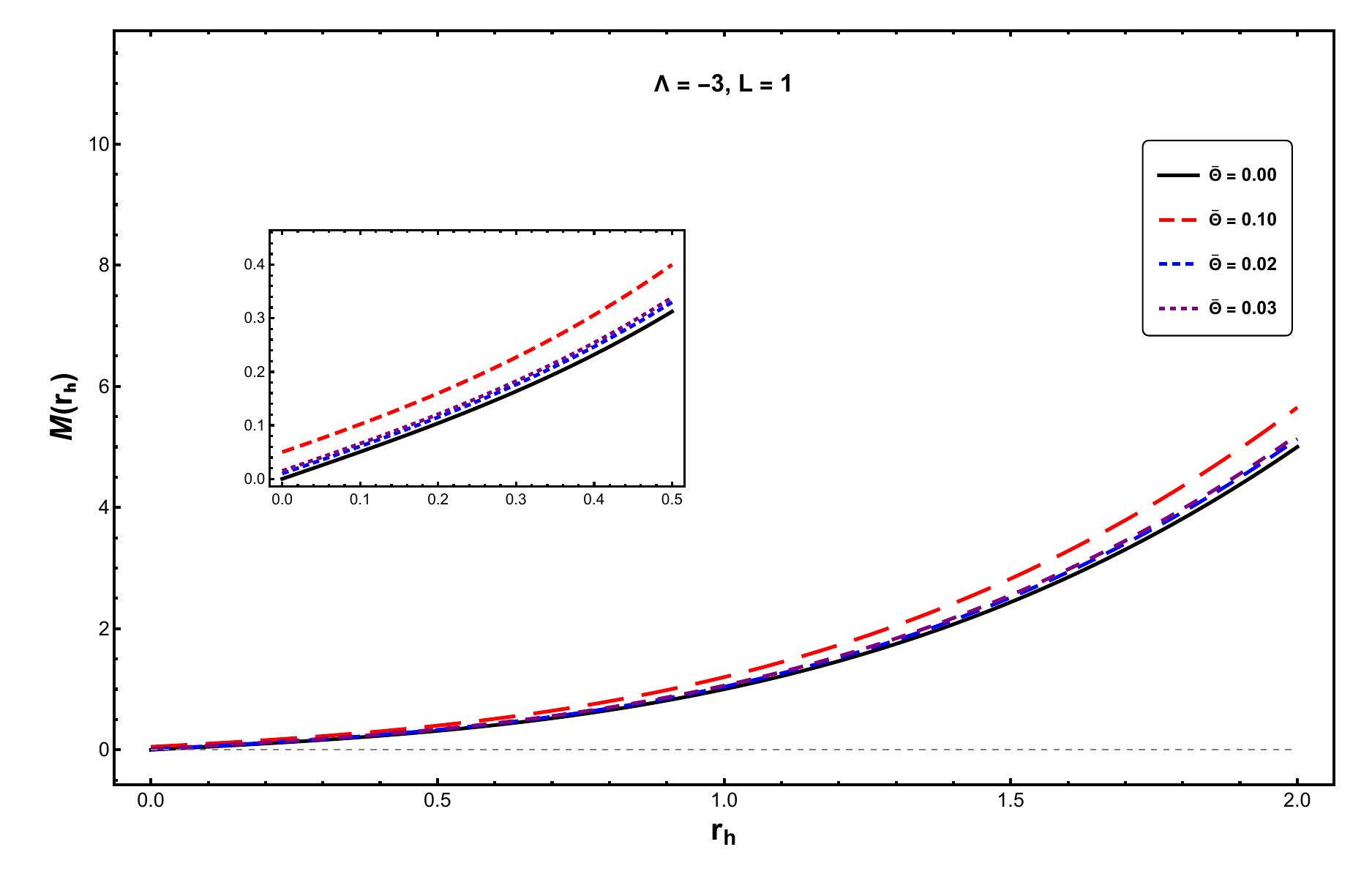}
\caption{The mass $M(r_h)$ of a Schwarzschild--AdS black hole in NC geometry, plotted for different NC parameters $\overline{\Theta}$, with $\Lambda=-3$ and $L=1$.}
\label{fig:M_rh}
\end{figure}

Figure~\ref{fig:M_rh} illustrates the behavior of the NC AdS black hole mass as a function of the horizon radius, according to Eq.~\eqref{eq17}. We observe the presence of a lower bound $M_{0}=\frac{\overline{\Theta}}{2}$, occurring at $\hat{r}_{h}=\frac{\overline{\Theta}}{2}$. Since $\overline{\Theta}$ is of the order of the Planck scale, this minimum mass is of the order of the Planck mass. It is clear that, in the NC framework, there is a minimum mass required to create a black hole. This supports the existence of a black hole remnant after Hawking evaporation due to noncommutativity effects, in agreement with previous results reported in Refs.~\cite{Alavi:2009tn, Wang:2024jlj}.

\section{Thermodynamics}

\subsection{Thermodynamic functions and the first law}

The NC temperature is defined using the NC surface gravity:
\begin{equation}\label{eq19}
\hat{T}=\frac{\hat{k}}{2\pi}.
\end{equation}
The metric~\eqref{eq7} is static, and is written in a manifestly time-independent form with vanishing time-space components $\hat{g}_{tj}=0$, then the surface gravity expression is given by~\cite{Magos:2020ykt}
\begin{equation}\label{eq20}
\hat{k}^{2}=-\frac{1}{4}\hat{g}^{tt}\hat{g}^{ij} \left(\partial_i\hat{g}_{tt}\right) \left(\partial_j\hat{g}_{tt}\right)=\frac{1}{4}\left(\partial_r\hat{g}_{00}\right)^2.
\end{equation}
Thus, Eq.~\eqref{eq19} reduces to
\begin{equation}\label{eq21}
\hat{T}=\frac{1}{4\pi}\frac{\partial \hat{g}_{00}}{\partial r}\bigg|_{r=\hat{r}_{h}}.
\end{equation}
Using the deformed metric components and the mass expression in Eq.~\eqref{eq17}, we obtain the NC temperature
\begin{equation}
\hat{T}=\frac{1}{4\pi r_{h}}\left(1-\Lambda r_{h}^{2}\right)-
\frac{\overline{\Theta} L}{4\pi r_{h}^{2}}\left(1+\Lambda r_{h}^{2}\right)+\mathcal{O}(\overline{\Theta}^{2}).
\label{eq22}
\end{equation}
The first term, $\frac{1}{4\pi r_{h}}\left(1-\Lambda r_{h}^{2}\right)$, corresponds to the temperature of an ordinary Schwarzschild--AdS black hole, while the second term, $-\frac{\overline{\Theta} L}{4\pi r_{h}^{2}}\left(1+\Lambda r_{h}^{2}\right)$, represents the NC correction.

\begin{figure}[ht]
    \centering
    \includegraphics[width=0.8\linewidth]{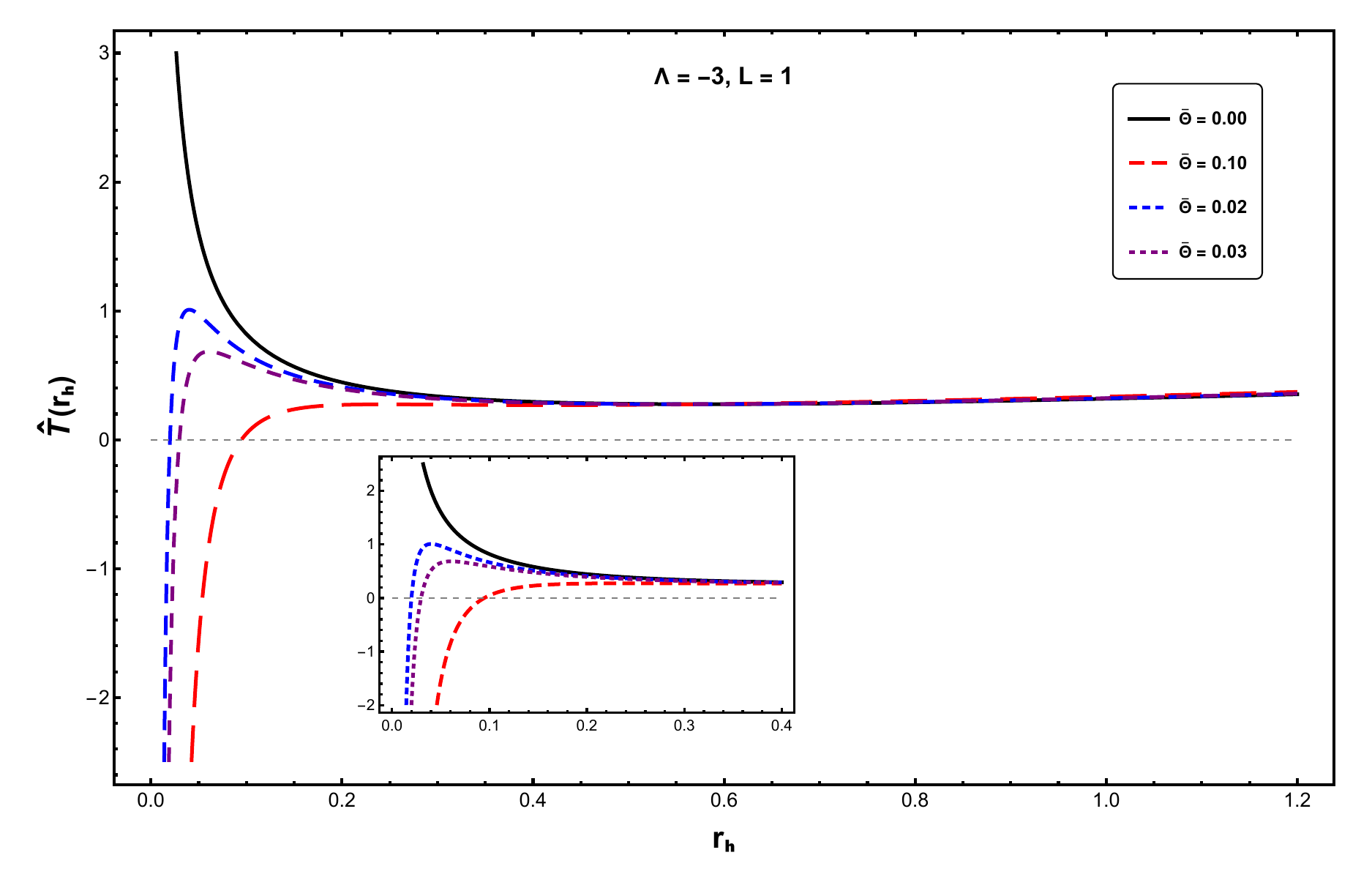}
    \caption{The Hawking temperature $\hat{T}(r_h)$ of a Schwarzschild--AdS black hole in NC geometry, plotted for different NC parameters $\overline{\Theta}$, with $\Lambda=-3$ and $L=1$.}
    \label{fig:T_rh}
\end{figure}

Figure~\ref{fig:T_rh} shows the behavior of $\hat{T}(r_{h})$ for a Schwarzschild--AdS black hole in NC geometry for various values of $\overline{\Theta}$. In the limit $r_{h} \to 0$, noncommutativity removes the divergence of the Hawking temperature, in analogy with the effect of electric charge in the Reissner--Nordstr\"om black hole~\cite{Kubiznak:2012wp}. As $r_{h}$ decreases, the surface temperature $\hat{T}$ increases and reaches a maximum value $\hat{T}^{\text{max}} = 0.02 / (\overline{\Theta} L)$ at the critical horizon radius $r_{h}^{c} = 2\overline{\Theta} L$. It then rapidly decreases to zero at the minimum horizon radius $r_{h}^{\text{min}} = \overline{\Theta} L$. This behavior resembles that of quantum-corrected black holes~\cite{Nozari:2006bi,Nicolini:2005vd}.

The thermodynamic behavior of a noncommutativity-corrected black hole is qualitatively similar to that of a Schwarzschild--AdS black hole with a conformal anomaly, with the main difference being the presence of a temperature divergence in the latter case~\cite{Cai:2014jea}. We observe that decreasing the NC parameter $\overline{\Theta}$ leads to a higher maximum surface temperature, indicating that $\overline{\Theta}$ acts as a quantum correction parameter. For $\overline{\Theta}=0$, $\hat{T}$ reduces to the ordinary Hawking temperature of a Schwarzschild--AdS black hole. The thermal energy can be identified as $E^{\text{th}} = \hat{T}^{\text{max}}$, and the corresponding black hole mass reads
\begin{equation}\label{eq23}
M = \frac{1}{G}\overline{\Theta}= \overline{\Theta} M_{\text{Planck}}^{2}.
\end{equation}

At the critical point $r_{h}^{c}$, the thermal energy and mass of the black hole are of the same order of magnitude. The NC parameter can then be estimated as
\begin{equation}\label{eq24}
\Theta\approx 0.1 \times \ell_p .
\end{equation}
This result shows that the noncommutativity parameter of spacetime is linked to the Planck scale $\ell_p$, consistent with experimental constraints from gravitational-wave data~\cite{Kobakhidze:2016cqh}, as well as with previous analyses of NC Schwarzschild black hole thermodynamics. Similar studies have also estimated $\Theta$ to be around $0.1 \times \ell_p$ by examining black hole thermodynamics~\cite{Campos:2021sff,Nozari:2006bi,Nicolini:2005vd,Wang:2024jlj}.

The entropy of the NC black hole is given by
\begin{equation}\label{eq25}
\hat{S} = \left. \frac{\hat{A}}{4} \right|_{r=\hat{r}_{h}},
\end{equation}
where $\hat{A}$ is the area of the event horizon in NC spacetime:
\begin{equation}\label{eq26}
\hat{A}=\int\int\sqrt{\hat{g}_{\theta\theta}\hat{g}_{\varphi\varphi}}\, d\theta\, d\varphi= 4\pi r_{h}^{2}\left(1 + \frac{2\overline{\Theta} L}{r_{h}}\right).
\end{equation}
Hence, the entropy of the AdS black hole in NC geometry reads
\begin{equation}\label{eq27}
\hat{S}(r_{h},\overline{\Theta}) = \pi r_{h}^{2}\left(1 + \frac{2\overline{\Theta}L}{r_{h}}\right)= S^{\text{o}} + 2\pi\overline{\Theta} L r_{h} + \mathcal{O}(\overline{\Theta}^{2}),
\end{equation}
where $S^{\text{o}} = \pi r_{h}^{2}$ is the entropy in the commutative case corresponding to $\overline{\Theta}=0$. The entropy thus increases by
\begin{equation}\label{eq28}
\hat{S} - S^{\text{o}} = 2\pi\overline{\Theta} L r_{h}.
\end{equation}

\begin{figure}[ht]
    \centering
    \includegraphics[width=0.8\linewidth]{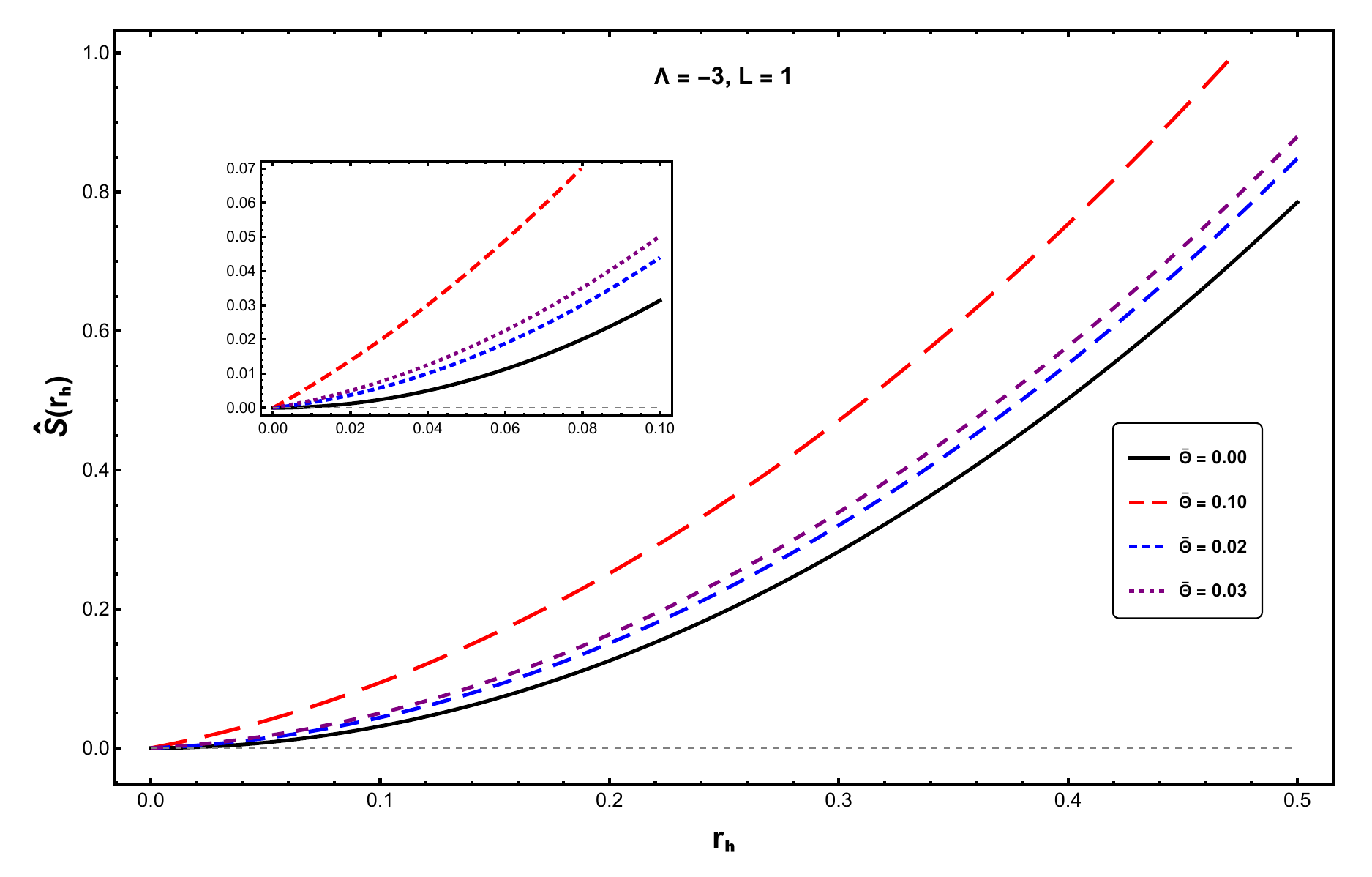}
    \caption{Entropy $\hat{S}(r_h)$ of a Schwarzschild--AdS black hole in NC geometry, plotted for different NC parameters $\overline{\Theta}$, with $\Lambda=-3$ and $L=1$.}
    \label{fig:S_rh}
\end{figure}
Figure~\ref{fig:S_rh} shows the entropy evolution as a function of the horizon radius. As $r_{h}$ increases, the entropy exhibits a clear deviation from the area law. For smaller $r_{h}$, this deviation gradually diminishes, and as $r_{h} \to 0$, the entropy vanishes, recovering the standard area law behavior.

In the thermodynamic interpretation of AdS black holes, the cosmological constant can be thought of as a perfect fluid stress-energy with pressure~\cite{Kastor:2009wy}
\begin{equation}\label{eq29}
P= -\frac{\Lambda}{8\pi} = \frac{3}{8\pi l^{2}},
\end{equation}
where $l$ is the AdS radius related to the cosmological constant by $\Lambda=-3/l^{2}$.

The NC black hole thus satisfies the first law of thermodynamics
\begin{equation}\label{eq30}
dM = \hat{T} d\hat{S} + V dP,
\end{equation}
where $V=\left(\frac{\partial M}{\partial P}\right)_{\hat{S}}$ is the corresponding thermodynamic volume of a NC black hole. It is given by
\begin{equation}\label{eq31}
V=\left(\frac{\partial M}{\partial P}\right)_{r_h}=\frac{4\pi}{3}r_h^{3}+4\pi\overline{\Theta} Lr_h^{2}.
\end{equation}
For a fixed cosmological constant, the second term in Eq.~\eqref{eq30} vanishes. From Eqs.~\eqref{eq17} and~\eqref{eq27}, we find
\begin{align}
\hat{T} &=\left( \frac{\partial M}{\partial r_{h}} \right)\left( \frac{\partial \hat{S}}{\partial r_{h}} \right)^{-1}\nonumber\\
&=\frac{1}{4\pi r_{h}}\left(1-\Lambda r_{h}^{2}\right)-\frac{\overline{\Theta} L}{4\pi r_{h}^{2}}\left(1+\Lambda r_{h}^{2}\right)+\mathcal{O}(\overline{\Theta}^{2}).\label{eq32}
\end{align}
The expression for $\hat{T}$ obtained here is identical to that derived from the surface gravity, confirming that the differential form $\left( \frac{\partial M}{\partial \hat{S}} \right)=\hat{T}$ holds. Therefore, the mass definition in Eq.~\eqref{eq17} satisfies the first law of thermodynamics, in contrast to previous claims~\cite{Wang:2024jlj,Singh:2020xju,Wang:2024jtp,Rodrigues:2022qdp} that it might be violated. 

The significance of the approach used in this study lies in the fact that the NC space defined by Eq.~\eqref{eq1} induces a coordinate displacement proportional to the NC parameter $\Theta^{\mu \nu}$. This displacement incorporates a constant vector that preserves Einstein's equation $(R_{\mu\nu} - \frac{1}{2}R g_{\mu\nu} - \Lambda g_{\mu\nu} = 0)$ to first order in $\Theta$, thereby keeping the stress-energy tensor zero. Consequently, the mass of the black hole in NC AdS space is reinterpreted as the NC enthalpy, or equivalently, the NC heat content, of spacetime~\cite{Kastor:2009wy,Kubiznak:2016qmn}. Furthermore, when the cosmological constant is absent $(\Lambda=0)$, the first law of black hole mechanics gives the well-known Smarr formula for static and asymptotically flat black holes.

It is easy to verify that the quantities $(M,\hat{T},\hat{S},P,V)$, given by Eqs.~\eqref{eq17}, \eqref{eq22}, \eqref{eq27}, \eqref{eq29}, and \eqref{eq31}, respectively, satisfy the Smarr relation
\begin{equation}\label{eq33}
M=2\left(\hat{T}\hat{S}-PV\right).
\end{equation}
We note that the thermodynamic quantities of the AdS black hole in NC space exhibit an interesting dependence on the noncommutativity parameter $\overline{\Theta}$. However, the noncommutativity parameter does not affect the first law of thermodynamics and the Smarr relation.

\subsection{Equation of state and critical point}

Using Eq.~\eqref{eq22} and the relation $P = -\frac{\Lambda}{8\pi}$, the pressure $P$ can be expressed as
\begin{equation}\label{eq34}
P=\frac{\hat{T}}{2r_h}\left(1-\frac{\overline{\Theta} L}{2r_h}\right)-\frac{1}{8\pi r_h^2}\left(1-\frac{\overline{\Theta} L}{r_h}\right).
\end{equation}
In order to relate the fluid volume to the horizon radius rather than the thermodynamic volume $V=\frac{4\pi}{3}r_h^3$, we first transform Eq.~\eqref{eq34} into a physical equation, where the physical pressure and temperature are given as follows
\begin{equation}
\begin{split}
P_{\rm Phys}&=\dfrac{\hbar c}{l_P^2}P, \\
\hat{T}_{\rm Phys}&=\dfrac{\hbar c}{k_B}\hat{T}.
\end{split}
\label{eq35}
\end{equation}
Then Eq.~\eqref{eq34} takes the following form
\begin{equation}\label{eq36}
P_{\rm Phys}=\frac{k_B\hat{T}_{\rm Phys}}{2l_P^2 r_h}+\cdots.
\end{equation}
To obtain an equation similar to the Van der Waals equation ($P=\frac{k_B T}{v}+\cdots$), we use the fact that the specific volume is $v=2l_P^2 r_h$. Returning to geometric units, we introduce the specific volume $v=2r_h$, which allows the equation of state to be rewritten as
\begin{align}
P&=\frac{\hat{T}}{v}\left(1-\frac{\overline{\Theta} L}{v}\right)-\frac{1}{2\pi v^2}\left(1-\frac{2\overline{\Theta} L}{v}\right)\nonumber\\
&=\frac{\hat{T}}{v}-\frac{1}{2\pi v^2}-\frac{\overline{\Theta} L\hat{T}}{v^2}+\frac{\overline{\Theta} L}{\pi v^3}.\label{eq37}
\end{align}

The critical point $(P_{c}, v_{c}, \hat{T}_{c})$ is determined from the inflection point conditions
\begin{equation}\label{eq38}
\left(\frac{\partial P}{\partial v}\right)_{\hat{T}}=\left(\frac{\partial^{2} P}{\partial v^{2}}\right)_{\hat{T}}=0 .
\end{equation}
Solving these conditions yields
\begin{equation}\label{eq39}
\frac{1}{\hat{T}_{c}} = \frac{2\pi\overline{\Theta} L}{2 - \sqrt{3}},\quad v_{c} = \left(3 + \sqrt{3}\right)\overline{\Theta} L, \quad \frac{1}{P_{c}} = \pi \left(3 + \sqrt{3}\right)^{3} \overline{\Theta}^{2} L^{2}.
\end{equation}
The critical ratio is then found to be
\begin{equation}\label{eq40}
\frac{P_{c}v_{c}}{\hat{T}_{c}} = \frac{1}{3} .
\end{equation}

It is worth noting that the value obtained here, $1/3$, is slightly smaller than the corresponding critical ratio $3/8$ predicted for a charged Schwarzschild--AdS black hole in the commutative case, which coincides with the well-known result for the Van der Waals fluid~\cite{Kubiznak:2012wp}. Interestingly, similar deviations from $3/8$ have also been reported for Schwarzschild--AdS black holes in loop quantum gravity~\cite{Wang:2024jtp}, as well as in NC geometry~\cite{Wang:2024jlj}, with shifts that are typically more pronounced. These deviations, which appear to be distributed around the canonical value $3/8$, may provide useful insight into the underlying thermodynamic structure of black holes and their correspondence with the equations of state of conventional fluids.

\begin{figure}[ht]
    \centering
    \includegraphics[width=0.8\linewidth]{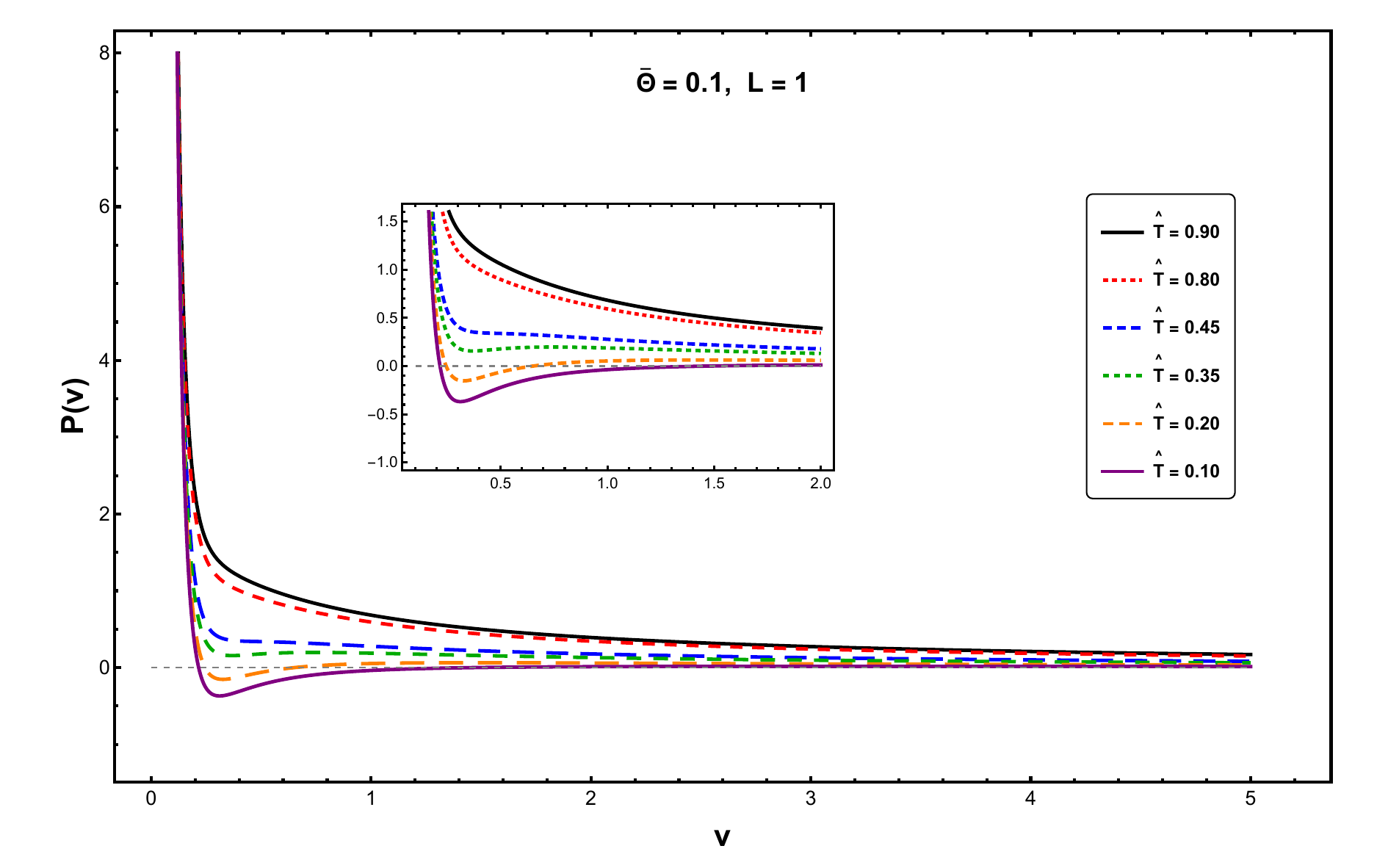}
    \caption{Pressure $P(v)$ of a Schwarzschild--AdS black hole in NC geometry, where $v$ denotes the specific volume, plotted for varying $\hat{T}$, with $\overline{\Theta}=0.1$ and $L=1$.}
    \label{fig:P_v}
\end{figure}

As shown in Fig.~\ref{fig:P_v}, for a fixed noncommutativity parameter $\overline{\Theta}$, the system exhibits a single critical point. The pressure--volume curves resemble those of a Van der Waals fluid, indicating the occurrence of phase transitions when $\hat{T}<\hat{T}_{c}$. In this regime, the system undergoes a first-order phase transition between small and large black holes. For $\hat{T}>\hat{T}_{c}$, the critical behavior disappears, and the pressure increases monotonically, corresponding to a single, large, and thermodynamically unstable black hole phase.

From Eq.~\eqref{eq34}, the NC equation of state can be recast in a more compact form
\begin{equation}\label{eq41}
\left(P+\frac{1}{2\pi v^{2}}\right)\left(v-2\overline{\Theta} L\right)=\hat{T}\left(1-\frac{3\overline{\Theta} L}{v}\right)=\tilde{T}.
\end{equation}

\subsection{Heat Capacity and Free Energy}

Another key quantity in the thermodynamic study of black holes is the specific heat capacity at constant pressure, $\hat{C}$ which characterizes the local thermodynamic stability of the black hole. It is related to the surface temperature and entropy through the relation
\begin{align}
\hat{C}\left(r_{h}\right)&=\hat{T}\left(\frac{\partial \hat{S}}{\partial \hat{T}}\right)=\hat{T}\left(\frac{\partial \hat{S}}{\partial r_{h}}\right)\left(\frac{\partial \hat{T}}{\partial r_{h}}\right)^{-1}\notag\\
&=2\pi r_h^2\frac{-\Lambda r_h^3+r_h-2\overline{\Theta}\Lambda L}{-\Lambda r_h^3-r_h+2\overline{\Theta}\Lambda L}.\label{eq42}
\end{align}
\begin{figure}[ht]
    \centering
    \begin{minipage}{0.48\linewidth}
        \centering
        \includegraphics[width=\linewidth]{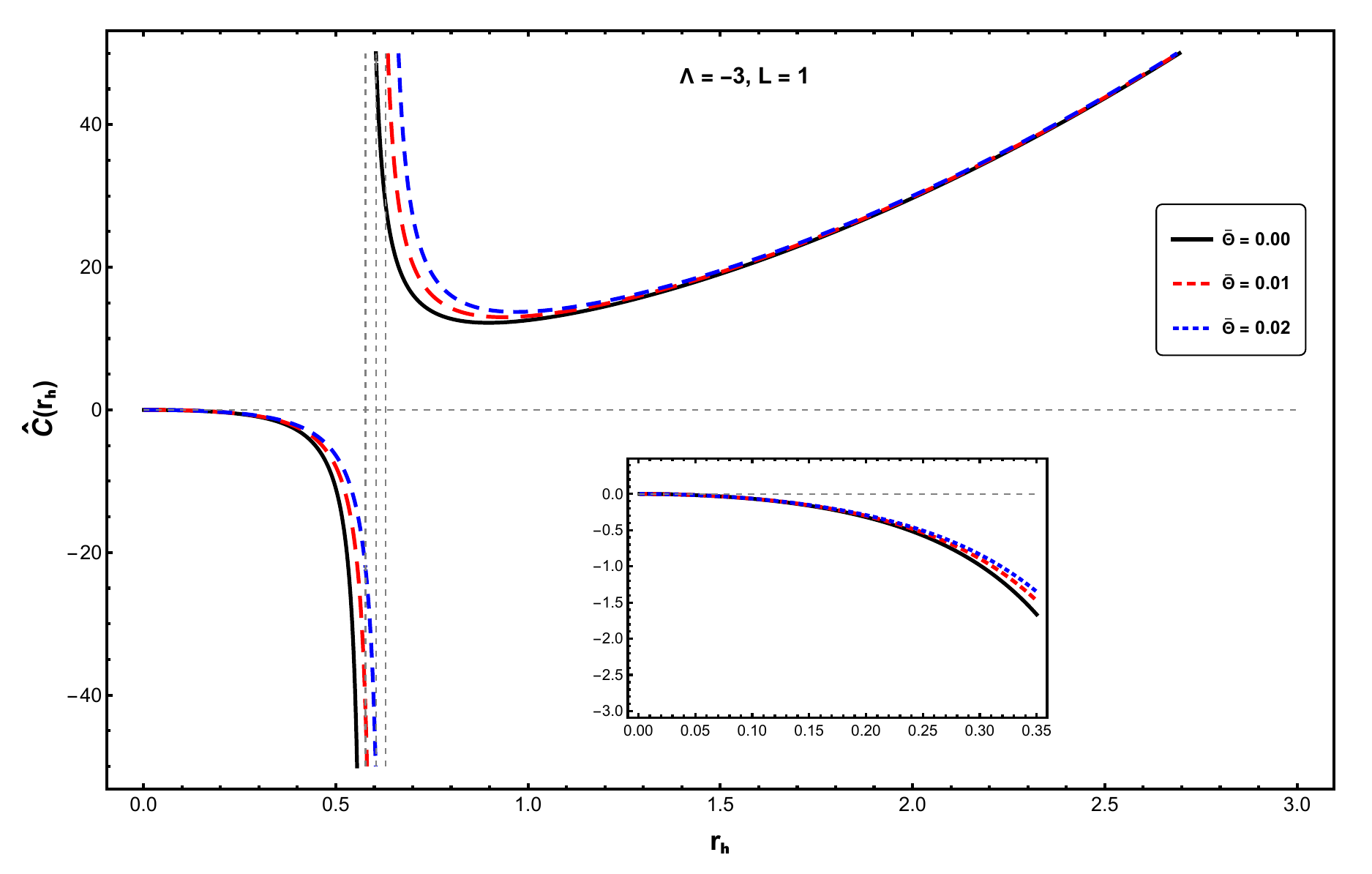}
    \end{minipage}
    \hfill
    \begin{minipage}{0.48\linewidth}
        \centering
        \includegraphics[width=\linewidth]{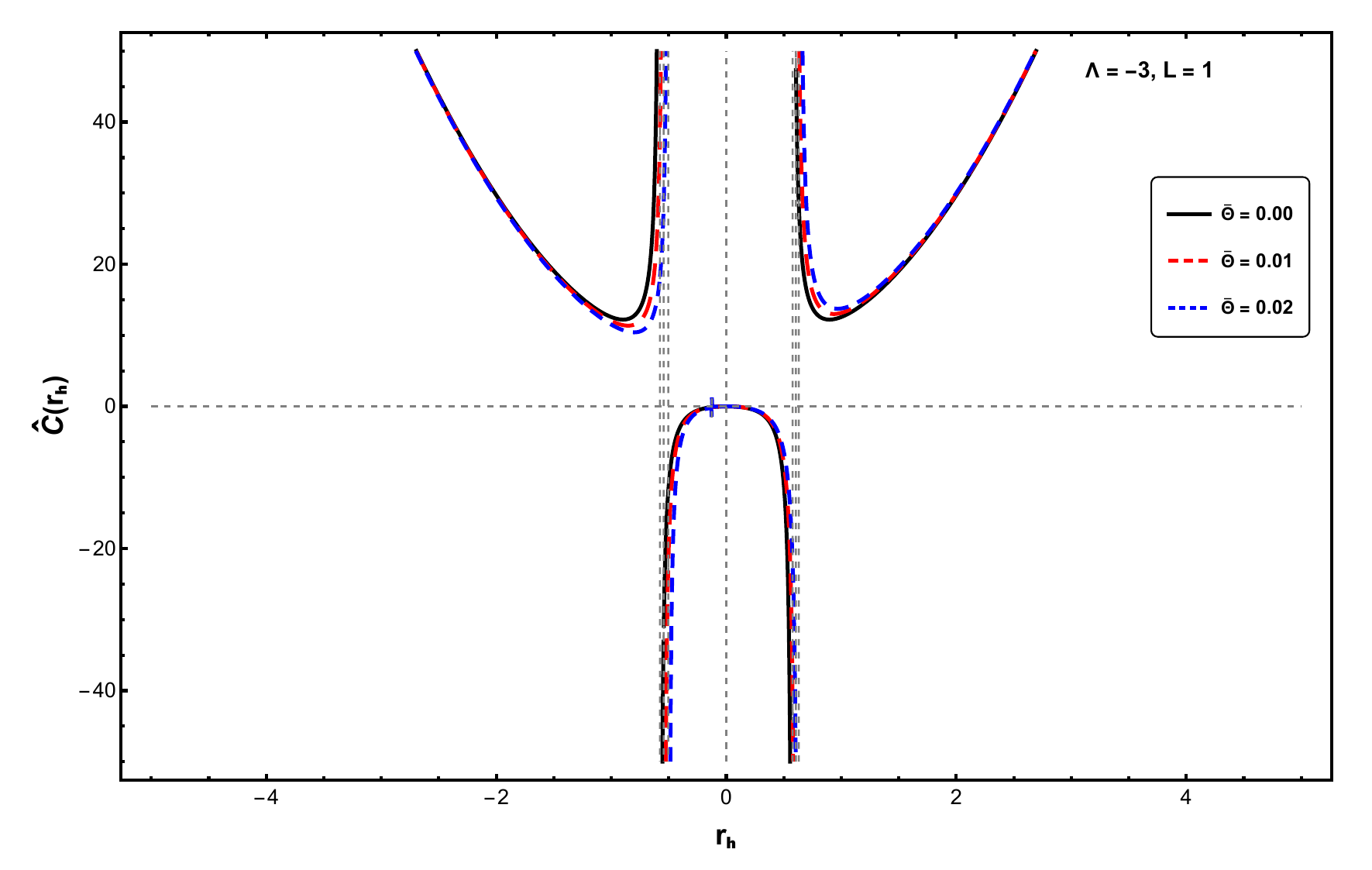}
    \end{minipage}
    \caption{The heat capacity $\hat{C}(r_h)$ of a Schwarzschild--AdS black hole in NC geometry, plotted for different NC parameters $\overline{\Theta}$, with $\Lambda=-3$ and $L=1$.}
    \label{fig:C_rh}
\end{figure}

Figure~\ref{fig:C_rh} illustrates the behavior of $\hat{C}(r_h)$ for fixed $\Lambda=-3$, $L=1$, and various values of $\overline{\Theta}$. The criterion used to determine whether a particular phase of a black hole is thermally stable relies primarily on its heat capacity. A positive heat capacity indicates a stable phase, while a negative heat capacity indicates an unstable phase. For $r_h<r_h^c$, the heat capacity is negative ($\hat{C}<0$), implying that the black hole is thermodynamically unstable and cannot reach local equilibrium. In contrast, for $r_h>r_h^c$, we have $\hat{C}>0$, indicating thermodynamic stability. When $\overline{\Theta}=0$, we recover the standard Schwarzschild--AdS behavior, where $\hat{C}$ changes sign at $r_h=r_h^c$, confirming the existence of a phase transition. Physically, this means that black holes with small event horizons are unstable, while those with large horizons are stable. Moreover, $\hat{C}\rightarrow0^{-}$ as $r_h\rightarrow0$, suggesting a physical limitation point~\cite{Sharif:2011ja}, while it diverges and changes sign at $r_h=r_h^c$, signaling a phase transition.
As $\overline{\Theta}$ increases, the radius of the locally stable black holes grows, and the corresponding value of $\hat{C}$ increases. This implies that stable black holes take longer to radiate and evaporate. In NC geometry, smaller black holes retain a negative heat capacity, making them unstable, while larger black holes have a positive heat capacity, thus making the system stable. Furthermore, the divergence of heat capacity at a critical point $r_h^c$ induces a phase transition for the black hole in NC space. In the commutative case, the ordinary heat capacity is always negative; therefore, there is no black-hole phase transition. This transition occurs due to the nonzero mass minimum caused by noncommutativity, which prevents singularity and creates a stable, cold state.

To further explore the phase transitions, we consider the thermodynamic potential represented by the free energy
\begin{align}
\hat{F}(r_h,\overline{\Theta})&=M-\hat{T}\hat{S}\notag\\
&=\frac{1}{4}r_h\left(1+\frac{\Lambda}{3}r_h^2\right)+\frac{\overline{\Theta} L}{4}\left(1+\Lambda r_h^2\right)+\mathcal{O}(\overline{\Theta}^{2}).\label{eq43}
\end{align}

\begin{figure}[ht]
    \centering
    \includegraphics[width=0.8\linewidth]{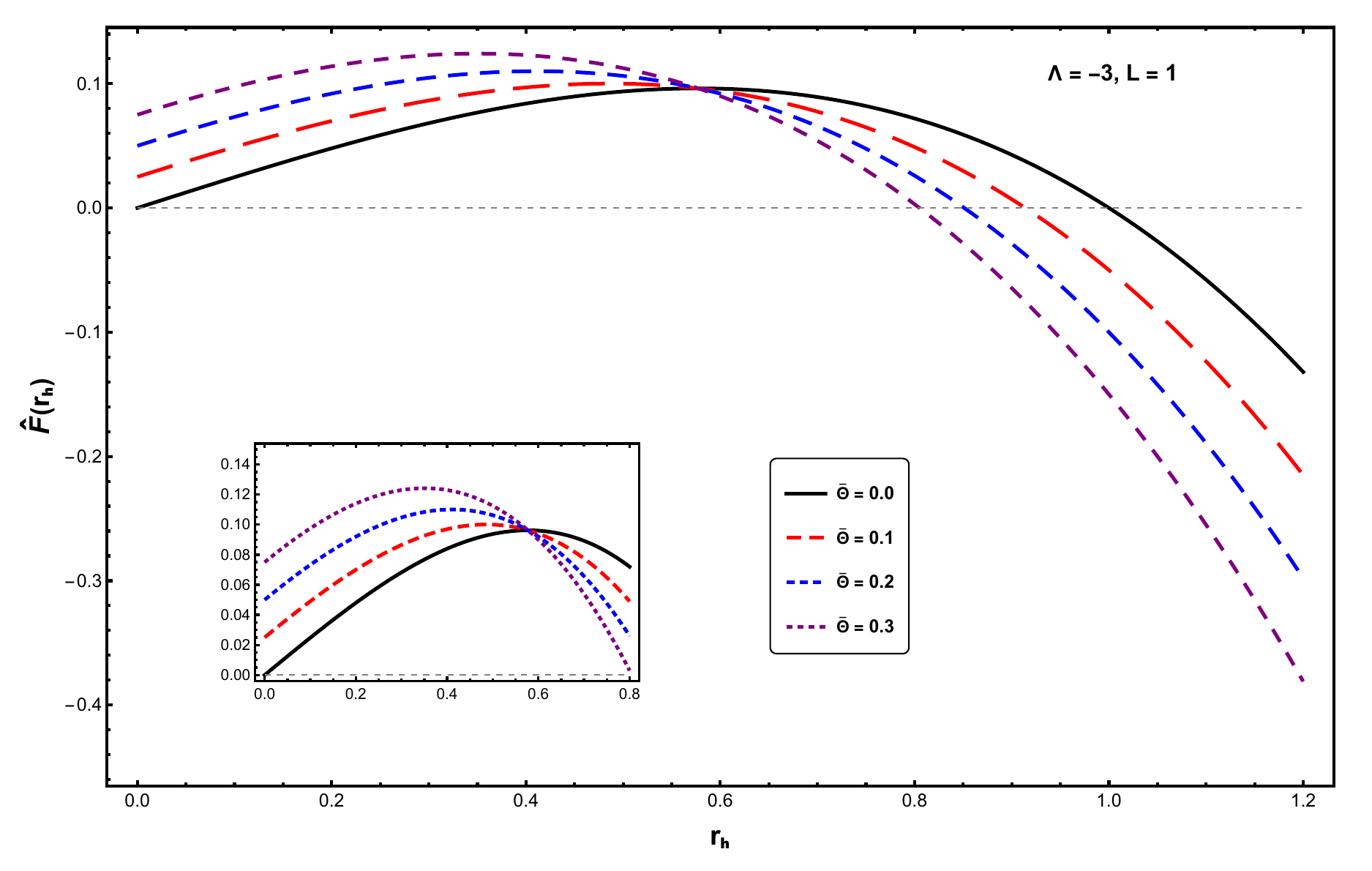}
    \caption{The free energy $\hat{F}(r_h)$ of a Schwarzschild--AdS black hole in NC geometry, plotted for different NC parameters $\overline{\Theta}$, with $\Lambda=-3$ and $L=1$.}
    \label{fig:F1_rh}
\end{figure}

Figure~\ref{fig:F1_rh} shows that the extrema of the free energy $\hat{F}$ are determined by
\begin{equation}\label{eq44}
\frac{\partial \hat{F}}{\partial r_{h}} = 0,
\end{equation}
where the location of this maximum shifts to smaller radii as $\overline{\Theta}$ increases. This region corresponds to instability, confirming the conclusion drawn from Fig.~\ref{fig:C_rh}: large black holes are thermodynamically stable, while small ones are not.

Using the expressions for the mass in Eq.~\eqref{eq17} and entropy in Eq.~\eqref{eq27} in NC Schwarzschild--AdS spacetime, the free energy can be expressed as
\begin{equation}\label{eq45}
\hat{F}(r_{h}, \hat{T}, \overline{\Theta}) = \frac{1}{2}r_{h}\left(1 - \frac{\Lambda}{3}r_{h}^{2}\right)- \pi \hat{T}r_{h}^{2} + \left(1 - \Lambda r_{h}^{2} - 4\pi \hat{T}r_{h}\right)\frac{\overline{\Theta} L}{2}.
\end{equation}
The extremum condition $\partial \hat{F}/\partial r_{h} = 0$ yields the corresponding Hawking temperature in NC Schwarzschild--AdS spacetime
\begin{equation}\label{eq46}
\hat{T}_{H} =\frac{1 - \Lambda r_{h}^{2} - 2\Lambda \overline{\Theta} L r_{h}}{4\pi r_{h} + 4\pi \overline{\Theta} L}\approx \frac{1 - \Lambda r_{h}^{2}}{4\pi r_{h}}-\frac{1 + \Lambda r_{h}^{2}}{4\pi r_{h}^{2}}\,\overline{\Theta} L.
\end{equation}

\begin{figure}[ht]
    \centering
    \includegraphics[width=0.8\linewidth]{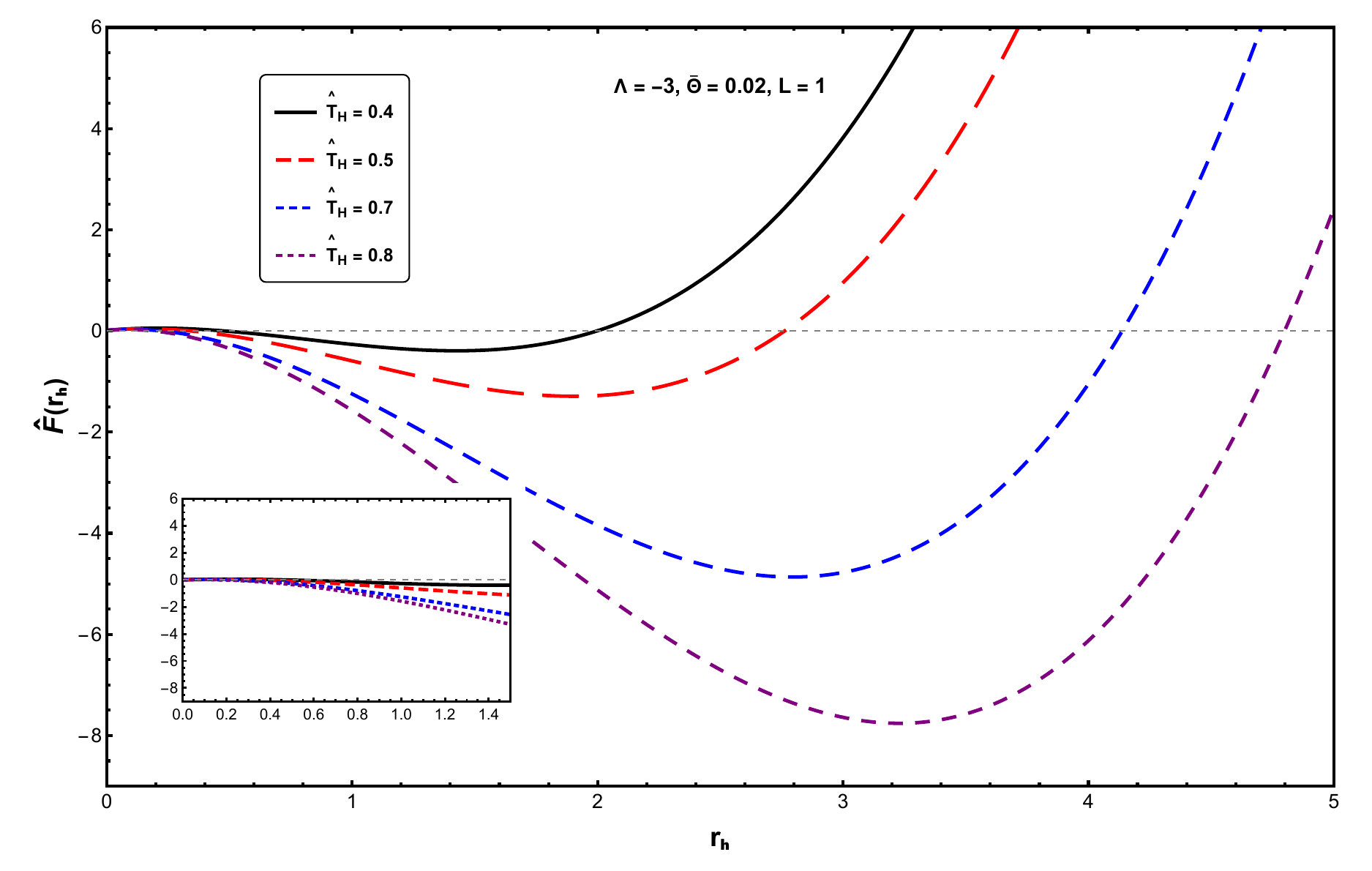}
    \caption{The free energy $\hat{F}(r_h)$ of a Schwarzschild--AdS black hole in NC geometry, plotted for different NC Hawking temperatures $\hat{T}_{H}$, with $\Lambda=-3$, $\overline{\Theta}=0.02$, and $L=1$.}
    \label{fig:F2_rh}
\end{figure}

As shown in Fig.~\ref{fig:F2_rh}, the maximum of $\hat{F}$ occurs in the region of smaller black holes, while the free energy profile changes as $\hat{T}_{H}$ increases.
In this region, $\hat{F} > 0$, indicating instability, whereas for larger $r_{h}$, $\hat{F}$ becomes negative, corresponding to stable configurations.  
When the temperature increases, $\hat{T}>\hat{T}_{H}$, for fixed $\Lambda$ and $\overline{\Theta}$, the free energy evolves from a positive maximum to a negative minimum; thus, the unstable small black hole evaporates into a stable, larger one. 
Thus, as $\hat{T}_{H}$ rises beyond its critical limit, the unstable phase contracts while the stable phase expands.

\section{Conclusions}

In this work, we have studied the phase transition and thermodynamic behavior of the SAdS-BH within the framework of NC geometry. By incorporating NC corrections to the SAdS spacetime metric up to first order in the NC parameter $\Theta$, as presented in~\cite{Larbi:2024rxv}, we derived the thermodynamic quantities of the NC SAdS-BH. The total mass, Hawking temperature, and entropy were obtained with first-order corrections in $\Theta$. 

We then examined the impact of noncommutativity on the first law of thermodynamics and verified that it remains exactly satisfied. We further showed that NC effects remove the divergence of the Hawking temperature present in the commutative case, replacing it with a finite maximum temperature attained during black hole evaporation, followed by a cooling phase down to zero temperature at a new minimum horizon radius associated with a minimum mass $M_{\min}$.

A detailed analysis of the free energy of the NC Schwarzschild--AdS black hole revealed two distinct extrema: a stable minimum corresponding to a large black hole phase and an unstable maximum associated with a smaller black hole. This structure is fully consistent with the behavior of the temperature profile (Fig.~\ref{fig:T_rh}) and the heat capacity analysis (Fig.~\ref{fig:C_rh}), both of which signal a phase transition between small and large black hole phases.

We also confirmed the existence of a phase transition in the NC AdS black hole system and uncovered a novel evaporation scenario leading to the formation of a microscopic residual black hole characterized by a minimum mass $M_{\min}=0.5\Theta$ and horizon radius $r_h=0.5\Theta$. This object may be interpreted as a black hole remnant, although it is found to be thermodynamically unstable. In addition, we showed that the equation of state of the NC Schwarzschild--AdS black hole exhibits a structure analogous to the Van der Waals fluid equation, pointing toward a deeper correspondence between NC black hole thermodynamics and conventional thermodynamic systems.

Finally, our analysis suggests a lower bound on the NC parameter of order $\Theta\sim 0.1\,\ell_P$, namely at the scale of the Planck length. Future work will focus on exploring higher-order NC corrections to Schwarzschild--AdS spacetime using methods from NC gauge theory.

\section*{Acknowledgements}

This work was supported by PRFU research project No. B00L02UN050120230003. The authors gratefully acknowledge financial support from the Algerian Ministry of Higher Education and Scientific Research and the Directorate General for Scientific Research and Technological Development (DGRSDT).

\bibliographystyle{unsrtnat}

\bibliography{Refs}

\end{document}